\documentclass[10pt,journal,compsoc]{IEEEtran}
\ifCLASSOPTIONcompsoc
  \usepackage[nocompress]{cite}
\else
  \usepackage{cite}
\fi
\ifCLASSINFOpdf
\else
\fi
\usepackage{xcolor}
\usepackage{comment}
\newcommand{\etal}{\textit{et al.}}

\usepackage{graphicx} 
\usepackage{float} 
\usepackage{subfigure} 
\usepackage{amsmath,bm}
\begin{document}
%
\title{Multi-User Redirected Walking in Separate Physical Spaces for Online VR Scenarios}
%
%
%
%

\author{Sen-Zhe~Xu,
        Jia-Hong~Liu,
        Miao~Wang,~\IEEEmembership{Member,~IEEE,}
        Fang-Lue~Zhang,~\IEEEmembership{Member,~IEEE,}
        and~Song-Hai~Zhang,~\IEEEmembership{Member,~IEEE}
\IEEEcompsocitemizethanks{\IEEEcompsocthanksitem Sen-Zhe Xu is with the YMSC, Tsinghua University, Beijing, China and BIMSA, Beijing, China.
\protect\\ E-mail: xsz@tsinghua.edu.cn
\IEEEcompsocthanksitem Jia-Hong Liu and Song-Hai Zhang are with the Department of Computer
Science and Technology, Tsinghua University, Beijing, China and BNRist, Tsinghua University, Beijing, China.
\protect\\ E-mail: liujiaho19@mails.tsinghua.edu.cn, shz@tsinghua.edu.cn
\IEEEcompsocthanksitem Miao Wang is with the School of Computer Science and Engineering, Beihang University, Beijing, China.
\protect\\ E-mail: miaow@buaa.edu.cn
\IEEEcompsocthanksitem Fang-Lue Zhang is with the School of Engineering and Computer Science, Victoria University of Wellington, Wellington, New Zealand.
\protect\\ E-mail: fanglue.zhang@vuw.ac.nz
\IEEEcompsocthanksitem Song-Hai Zhang is the corresponding author.
}
\thanks{Manuscript received -, 2022; revised -, 2022.}
}

%
%

\markboth{Journal of \LaTeX\ Class Files,~Vol.~14, No.~8, August~2015}%
{Shell \MakeLowercase{\textit{et al.}}: Bare Demo of IEEEtran.cls for Computer Society Journals}
\IEEEtitleabstractindextext{%
\begin{abstract}
With the recent rise of Metaverse, online multiplayer VR applications are becoming increasingly prevalent worldwide. Allowing users to move easily in virtual environments is crucial for high-quality experiences in such collaborative VR applications. This paper focuses on redirected walking technology (RDW) to allow users to move beyond the confines of the limited physical environments (PE). The existing RDW methods lack the scheme to coordinate multiple users in different PEs, and thus have the issue of triggering too many resets for all the users. We propose a novel multi-user RDW method that is able to significantly reduce the overall reset number and give users a better immersive experience by providing a more continuous exploration. Our key idea is to first find out the "bottleneck" user that may cause all users to be reset and estimate the time to reset, and then redirect all the users to favorable poses during that maximized bottleneck time to ensure the subsequent resets can be postponed as much as possible. More particularly, we develop methods to estimate the time of possibly encountering obstacles and the reachable area for a specific pose to enable the prediction of the next reset caused by any user. Our experiments and user study found that our method outperforms existing RDW methods in online VR applications.
\end{abstract}

}

%
\maketitle

\IEEEdisplaynontitleabstractindextext

%
\IEEEpeerreviewmaketitle

\IEEEraisesectionheading{\section{Introduction}\label{sec:introduction}}

\begin{figure*}[h]
\centering
\includegraphics[width=\textwidth]{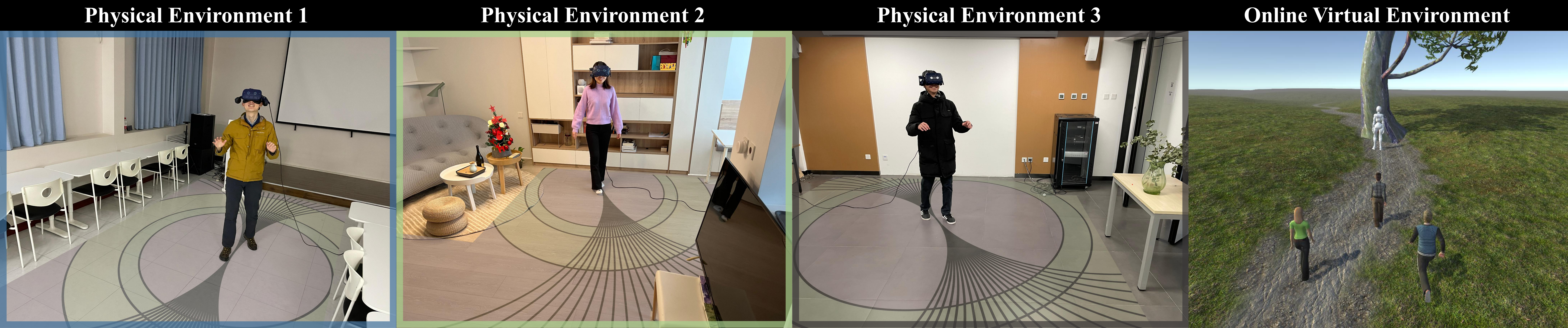}
\caption{Demonstration of multiple users in separate physical environments (PE), playing the same online VR game ``Red Light, Green Light" in an online virtual environment (VE). For the fairness of locomotion, all users need to be paused and reset when any of the users encounter obstacles. Our redirected walking controller analyses the time of encountering obstacles and spatial extent of reachable area for each user in real-time and significantly reduces the number of resets for all the users by rigorously reasoning and maximizing the "bottleneck" time to the subsequent reset.}
\label{fig:teaser}
\end{figure*}

Facilitated by the convergence between Internet technologies and Virtual Reality (VR), multiplayer online VR applications are becoming a reality. The recent rise of Metaverse~\cite{ondrejka2004escaping} has drawn the increasing interest of the Computer Graphics community to such kinds of immersive, interactive, and collaborative VR experiences. 
In a multiplayer online VR application/game, players are located in a separate physical space (e.g., their own home), but they can interact and play in a shared online virtual environment through the internet connection, as an example shown in Fig.~\ref{fig:teaser}. 

For offering VR users easy movements in the virtual environment (VE), it is widely accepted that natural walking provides a much higher presence for users~\cite{usoh1999walking,razzaque2001redirected,sakono2021redirected,williams2021arc}, compared with using controllers or Omni-directional treadmills for locomotion. However, during natural walking, space limitation of the physical environment (PE) often limits the user's walking in a large VE. Redirected walking (RDW)~\cite{razzaque2001redirected} has been proposed to reduce the number of encounters with physical obstacles and enable users to explore larger VEs in relatively small PEs. The RDW methods usually use \emph{subtle} and \emph{overt} manners together to adjust the user's movement. The \emph{subtle} RDW methods exploit the imprecision of the human motion perception and slightly modify the user's orientation or velocity during walking, enabling users to walk longer without noticing the inconsistency between their virtual and physical movements. Once the user is about to encounter an obstacle, the \emph{overt} RDW technology then will suspend the virtual scene and reset the user's orientation in PE. The previous RDW methods aim to reduce the number of resets since resets cause pauses in the user's immersive experience.

However, the previous RDW algorithms only focused on the situation of a single VR user or multiple VR users in a shared physical space. Although previous RDW methods for multiple users can avoid collisions among users, each user's resets are considered irrelevant to others. Since multiple users are located in different physical environments in online VR apps/games, a fairness issue emerges when applying the previous RDW algorithms to multiplayer online VR apps/games. When one of the users has to be reset due to the physical obstacles, the other users can continue to walk in their PEs, making the users' relative positions in the VE change. As the users' tasks for an online VR application/game are often correlated, the users' relative virtual position changes will cause intolerable unfairness in online VR apps/games. For example, real-time strategy (RTS) games such as ``Dota" or ``Red Light, Green Light" for online VR will become unplayable due to users' relative position changes. Furthermore, since users' PEs are different, users within smaller PEs will have fewer opportunities for moving during the online game, as they will be frequently suspended and reset their physical poses. For the fairness of the online VR apps/games, a good online RDW strategy must make the locomotion opportunities of different users equal, regardless of different physical environment layouts.

To achieve the locomotion fairness mentioned above in off-site multiplayer online VR apps/games, when a user gets stuck in the virtual space due to resetting, other users should not move in the virtual space either. A reasonable solution is that when some of the users trigger a reset, all users pause and reset to their better orientations synchronously and then resume the online game together. Rather than allowing the users to continue walking in PE while staying still in VE, this manner is less noticeable and more understandable to the users. More importantly, it can avoid wasting physical space. We refer to the manner that all players reset and resume together as the \textbf{locomotion fairness constraint}. Obviously, the \textbf{locomotion fairness constraint} would still increase the number of resets for every single user since other players may implicate each player to reset. Thus, we aim to reduce the number of resets of all the users for off-site multiplayer online VR apps/games under the \textbf{locomotion fairness constraint}.

In this paper, we propose the first practical and reliable RDW solution for off-site multiplayer online VR apps/games, which takes into account the \textbf{locomotion fairness constraint} and keeps the relative virtual positions unchanged after resets, as well as reduces the number of resets for all the connected users.
Our key idea is to coordinates all the users' walking to maximize the time intervals between resets. Specifically, we propose practical yet effective methods to estimate the time range of encountering obstacles for each user's pose and the spatial extent achievable in a given time after resetting. By analyzing how long and how far users can walk after a reset we find the ``bottleneck" user that may cause all users to the next reset. The reset occurrence will be put off by making the bottleneck user walk for the longest possible time. To prolong the time before the further subsequent reset, the users other than the bottleneck user will be reset and redirected to the their reachable and favorable poses at the same time, so all the user can have the longer possible walking distances/times after the next reset caused by the bottleneck user. 

We conduct experiments in both simulation and real-world environments. We implement our work on the top of the open-source OpenRDW platform~\cite{li2021openrdw}, and by our work we have extended the OpenRDW platform to a ``Meta" version and it's also the first RDW platform for the off-site multiplayer online VR applications. Our code will be released. The results of both simulations and real-world experiments demonstrate that our method outperforms all compared techniques.

\section{Related Work}

\subsection{Redirected Walking Strategies}
The formulation of the RDW problem can be traced back to the year 2001\cite{razzaque2001redirected}. Many practical and far-reaching heuristic methods have been proposed since then. Razzaque \etal~\cite{razzaque2005redirected} proposed three general and widely used redirecting methods, namely, steer-to-center (S2C), steer-to-orbit (S2O), and steer-to-multiple-targets (S2MT), which respectively redirects the user to the physical center, the circular orbit around the physical center and the predefined waypoints in physical space. Hodgson \etal~\cite{hodgson2013comparing} improved S2C by setting temporary steering targets to make the user back to the tracking space center faster, which involves a new strategy named steer-to-multiple+center that combines S2C and S2MT by including physical centers to the waypoints in S2MT. Azmandian \etal~\cite{azmandian2015physical} compared various RDW methods in different physical environment sizes and aspect ratios and confirmed that S2C outperforms most other methods, but S2O performs better when the user walks along a long straight virtual path. 

The recent efforts on RDW are seeking more efficient solutions for redirecting users. The methods based on Artificial Potential Functions (APF) are influential among them. Thomas \etal's APF (TAPF)~\cite{thomas2019general} proposed Push/Pull Reactive, which utilizes the attractive or repulsive force calculated by the predefined APFs to push users away from obstacles or pull users towards desirable areas. Bachmann \etal~\cite{bachmann2019multi} use APF to repel the user from obstacles and other users, allowing multiple users to walk in the same tracking space. Messinger \etal~\cite{messinger2019effects} proposed a follow-up work of Bachmann \etal's APF, which refined the original work in the calculation of the force vectors and extended its adaptability in irregular tracking space. Dong \etal proposed Dynamic APF (DAPF)~\cite{dong2020dynamic}, which also uses the dynamic artificial potential field to deal with multi-user scenarios, which takes into account the joint repulsive forces from obstacles, the users, and the user's future state, as well as the gravitational force of the steering target.

Reinforcement learning-based methods are another rapidly raised branch of RDW methods. Lee \etal~\cite{lee2019real} proposed Steer-to-Optimal-Target (S2OT), which estimate optimal steering target from $25$ predefined targets through reinforcement learning. Lee \etal~\cite{lee2020optimal} later proposed Multiuser-Steer-to-Optimal-Target (MS2OT), a reinforcement learning-based RDW method for multiple users, which improves S2OT by considering pre-reset actions, using more steering targets, and improving the reward function. Strauss \etal~\cite{strauss2020steering} proposed Steering via Reinforcement Learning (SRL), which applies reinforcement learning technique to determine the locomotion gains when steering the users directly. Chang \etal~\cite{chang2019redirection} also proposed a method with plannability/versatility to redirect users using reinforcement learning. 

Recently alignment-based RDW methods~\cite{williams2021arc,williams2021redirected} have been proposed to redirect users. Thomas \etal~\cite{thomas2020towards} introduced Reactive Environmental Alignment (REA), which transits the physical and virtual environments from misaligned states to aligned states. Williams \etal~\cite{williams2021arc} proposed an alignment-based Redirection Controller (ARC) to make proximity to obstacles in PE matches that in VE. Williams \etal~\cite{williams2021redirected} also proposed to use visibility polygons to calculate the walkable spaces and align the visibility polygons of PE and VE with steering the user. Nevertheless, those algorithms perform alignment concerning obstacles in VE. Those algorithms will be less effective if VE contains no obstacles and the VE is infinite in size.

RDW methods for dynamic environments and multiplayer are also a research hotspot. Besides the above methods involving multi-users, other works are dedicated to redirecting multiple users. Bachmann \etal~\cite{bachmann2013collision} extended RDW controllers to support two-player VR apps by forecasting potential collisions. Azmandian \etal~\cite{azmandian2017evaluation} explored three fundamental components of collision prevention for two-user VR scenarios and compared collision prevention strategies with PE subdividing. Z.Dong\etal~\cite{dong2019redirected} proposed to redirect multiple users by smoothly mapping virtual scenes to PEs and avoiding collisions with dynamic virtual avatars. T.Dong \etal~\cite{dong2021dynamic} proposed an RDW method for multi-user navigation in large-scale virtual environments based on dynamic density. 

However, the existing RDW methods are primarily for single or multiple users in the shared physical space with irrelevant locomotion in VE. Effective RDW strategy for multi-user in separated PEs while related in a shared online VE with locomotion fairness is still a challenge. 

\subsection{Locomotion gains for Redirected Walking}
To manipulate the mapping between the virtual and real-world imperceptibly, certain gains should be applied. Early methods proposed to use scaled translation~\cite{williams2006updating} and augmenting motion in the user's intended direction~\cite{interrante2007seven} for redirecting the user. Steinicke \etal~\cite{steinicke2008taxonomy} then presented a taxonomy of redirection gains, including \emph{translation}, \emph{rotation} and \emph{curvature} gains. \emph{Translation} gains scale the user's positional displacement, making the user move faster or slower in the physical environment. \emph{Rotation} gains apply a scale factor to the user's rotating speed, leading to a faster or slower turn. \emph{Curvature} gains steer the user's direction when the user is walking forward. To avoid simulator sickness, the gains should be constrained within certain thresholds. It has been measured that in average case, translation gains should be in the range of $0.86$ to $1.26$, and rotation gains should be in the range of $0.8$ to $1.49$~\cite{steinicke2009real,steinicke2009estimation}. It has been measured and widely accepted by the recent works that the curvature gain should have a minimal radius of $7.5$m~\cite{azmandian2015physical,bachmann2013collision,hodgson2013comparing,thomas2019general,lee2020optimal,bachmann2019multi,messinger2019effects,dong2020dynamic,dong2021dynamic,williams2021arc,williams2021redirected}.

\section{Method}
Suppose we have $n$ users $U=\left\{u_1,u_2,...,u_n\right\}$ participating the same online VR application in $n$ separated physical spaces $\left\{E_1,E_2,...,E_n\right\}$. Without loss of generality, we assign each user to a mission for each trial, where a user $u_i$ has to move towards a virtual target located in the VE continuously. When $u_i$ reaches the target, a new virtual target for the user will appear to keep $u_i$ moving. We denote the latest target of the user $u_i$ as $w_i$. The targets $W=\left\{w_1,w_2,...,w_n\right\}$ for all the users can be at different locations in the VE. Since the system specifies the users' present tasks, the RDW controller knows the current $W$. In the interests of fairness, all users need to be frozen and reset when some of them encounter obstacles in their physical spaces to ensure the relative positions of all the users in VE remain unchanged. Thus the number of resets will increase for every single user compared with the single-player game, as the users are implicated in resetting even when they do not encounter a obstacle. Therefore, our goal is to minimize the overall number of resets for all the involved off-site users. Before we introduce our redirection strategy for all the users, we first start by describing the estimation of \emph{time before encountering obstacles} and \emph{spatial extent of the reachable area with a given time} for a single user in the following sections.

\subsection{Time Span before Encountering Obstacles}
\label{3.1}
In this subsection, we aim to figure out the time that a user could walk before encountering obstacles in PE when walking towards a target in VE. Given a user's current pose $P=\left(\boldsymbol{p},\theta\right)$, whereas $\boldsymbol{p}=\left(p_x,p_y\right)$ is the user's position and $\theta$ is the orientation, the user can be steered to different physical positions imperceptibly with different curvature gains. 
The user's motion typically consists of a series of segments of straight walk toward the current virtual target. For one segment of walk, it begins when the user turns around, or a reset ends and finishes when the user reaches the target or triggers a reset. We choose to steer the user with a constant curvature gain during one segment of their walk for users' comfort. To resolve the time that could be taken to encounter obstacles, we propose to use an \emph{inscribed circle strategy}. The idea is to scan the physical space with a series of radial curves from $P$, and estimate the time needed to hit an obstacle along these curves.
\begin{figure}[h]
    \centering
    \includegraphics[width=0.65\linewidth]{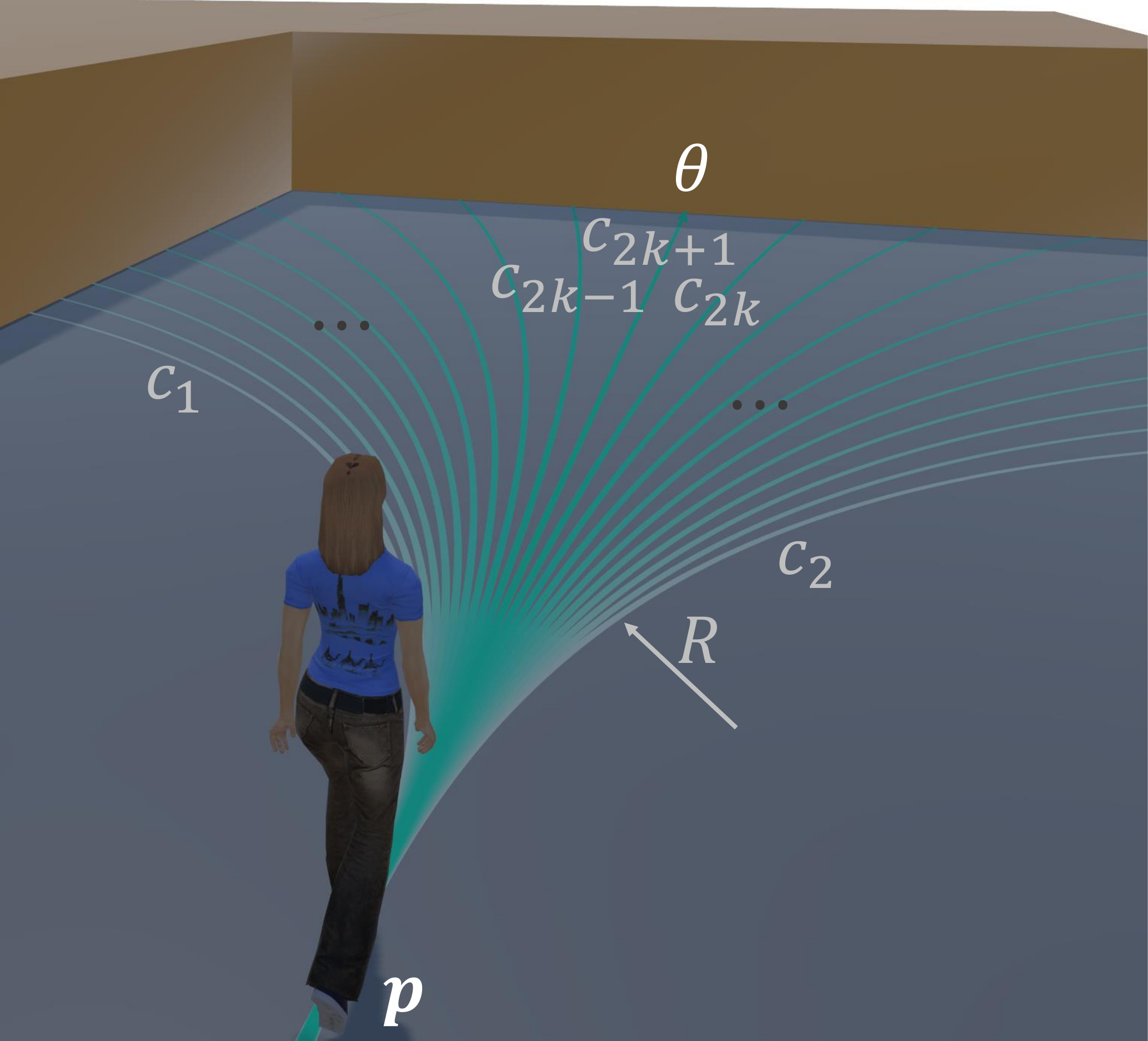}
    \caption{
    The \emph{inscribed circle strategy} for time span estimation of encountering obstacles.
    }
    \label{fig:3_1}
\end{figure}

Taking $\theta$ as the tangent direction and $\boldsymbol{p}$ as the contact point, we can draw a series of mutually inscribed circles $C=\left\{c_{i}\right\}$ to the left and right of $P$ with different radius $r_i=\lvert\alpha_i\rvert R$, as shown in Fig.~\ref{fig:3_1}. The equation of $c_i$ can be written as:
\begin{equation}
\label{eq:c_i}
\left\|
\begin{bmatrix} \cos\theta & \sin\theta \\ -\sin\theta & \cos\theta \end{bmatrix}
\begin{bmatrix} x-p_x \\ y-p_y      \end{bmatrix}
-
\begin{bmatrix} 0     \\ \alpha_i R \end{bmatrix} 
\right\|_2^2
=
\left(\alpha_i R \right)^2
\end{equation}

$R$ is the radius of the curvature gain threshold and $\lvert\alpha_i\rvert\geq 1$ is a factor for generating different curvature gains. Note that we can steer the user to the left or right of $\boldsymbol{p}$, the sign of $\alpha_i$ in equation~(\ref{eq:c_i}) just reflects the direction to steer, i.e., $\alpha_i>0$ means steering the user to the left, and $\alpha_i<0$ means right. If we steer the user using one of the $r_i\geq R$ as the radius of the curvature gain with the direction indicated by $\alpha_i$, the user's physical path will be coincident with the circle $c_i$. 
If $\alpha$ has $k$ pairs of opposite finite values, it is easy to know that there are a total of $2k$ circles starting from $\boldsymbol{p}$. To enable $C$ to scan the PE uniformly, we make $\alpha_i$ grow slowly when $i$ is close to $1$, and diverge to infinity as $i$ increases to $2k$. Specifically, we make our $\alpha_i$ as:
\begin{equation}
\alpha_i = \frac{\left(-1\right)^{\left(i-1\right)}}{\cos{\left(\frac{\pi}{2k}\times\lfloor \frac{i-1}{2} \rfloor\right)}}\quad \left(1\leq i \leq2k\right)
\end{equation}

We take circle $c_i$ as the candidate path and intersect it with all potential obstacles in PE, and get the length $l_i$ from $\boldsymbol{p}$ to the nearest obstacle along $c_i$. If $c_i$ happens to have no intersection with all obstacles (this rarely occurs unless the PE is extremely large), we designate $l_i$ as infinity which means the user can walk any length along this path. Suppose the user's walking speed in the virtual space is $v$ ($v$ can be different for different users), the permitted time for a user walking along path $c_i$ without considering translation gain is: 
\begin{equation}
t_i = \frac{l_i}{v}\quad \left(1\leq i \leq2k\right)
\end{equation}

It is worth noting that when $\alpha \rightarrow \infty$, the circle will approach a straight line, and it is equivalent to using no curvature gain when steering the user. We consider the straight path as $c_{2k+1}$ and calculate its intersection with the obstacles to obtain the walking time with no curvature gain as $t_{2k+1}$. Considering that we can also speed up or slow down a user's physical speed with a translation gain $g_t$, the possible range of walking time $t_i$ along $c_i$ is further scaled to $\left[{t_i}{g_t^{min}},{t_i}{g_t^{max}}\right]$. As long as $k$ is large to make the candidate paths $C$ sufficiently dense, we can approximate the maximum time that the user could have before encountering obstacles from a starting pose $P$ as:
\begin{equation}
T_{max}\left(P\right)=\max \limits_{i}{\left({t_i}{g_t^{max}}\right)}
\end{equation}

Considering that the user can voluntarily reset before encountering the obstacles, the RDW controller can make the user to reset at any time between $t\in\left[ 0,T_{max}\left(P\right) \right]$ when the user is starting from pose $P$.

\subsection{Spatial Extent of Reachable Area with Given Time}
\label{3.2}

When the user is about to reset, predicting the reachable area in time $T$ after that reset benefits our RDW controller to reset and redirect the users to more favorable positions and thus minimize the common number of resets. We then solve the problem of the spatial range of the positions that a user can be possibly navigated to within a given walking time $T$ supposing a reset just happened at a position $\boldsymbol{p}$ for the user.
\begin{figure}[h]
    \centering
    \includegraphics[width=1.0\linewidth]{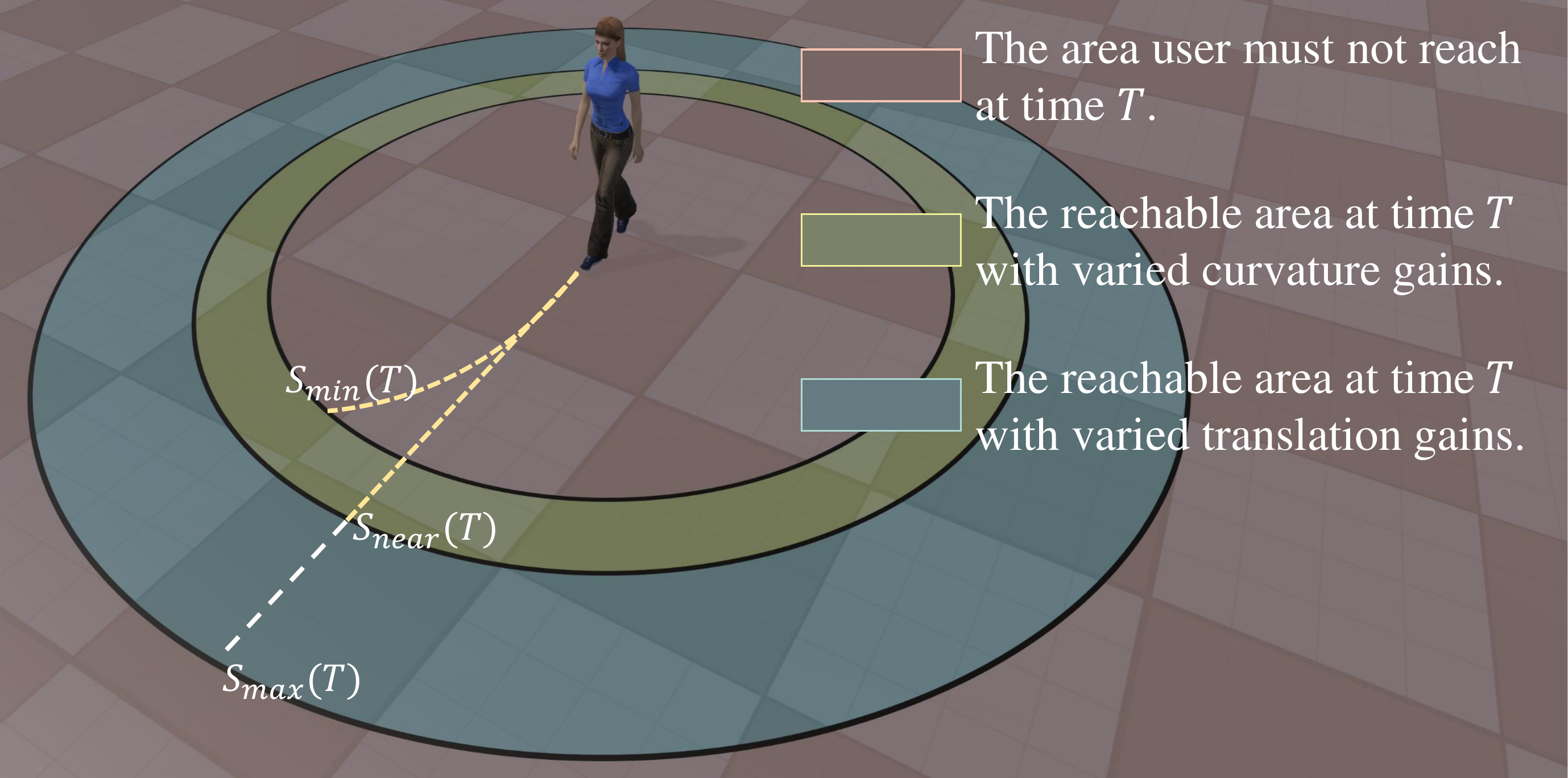}
    \caption{
    The \emph{concentric circles strategy} for spatial extent estimation of the reachable area within given time $T$.
    }
    \label{fig:3_2}
\end{figure}

We propose an intuitive and effective way called \emph{concentric circles strategy} to solve this problem. We found that since a user can face any needed direction after a reset, all their reachable positions must fall in a particular circle area centered at the user's current position $\boldsymbol{p}$ without considering the obstacles. We thus inspect the displacement $s$ from a position to the user's current position $\boldsymbol{p}$ to determine whether it is reachable for the user in time $T$ after a reset.

Let us consider the upper bound of the possible displacement $S_{max}$ first. To make the user walk to the position with the farthest displacement, $S_{max}$, the controller has to apply the maximum translation gain $g_{t}^{max}$ to make their walking speed faster. Meanwhile, the user must be kept to walk straight, and the controller should give no curvature gain; otherwise, their trajectory will be bent. Therefore, the maximum possible displacement for the user within time $T$ can be calculated as:
\begin{equation}
S_{max}\left(T\right) = \frac{v\cdot T}{g_{t}^{min}}
\end{equation}

Where $v$ is the user's speed in virtual space. Likewise, we can make the user walk for a shorter distance within the given time $T$ by simply adjusting the translation gain to the minimum threshold $g_{t}^{min}$:
\begin{equation}
S_{near}\left(T\right) = \frac{v\cdot T}{g_{t}^{max}}
\end{equation}

Conceivably, any position with a displacement of $s \in$ $[S_{near}\left(T\right),$ $S_{max}\left(T\right)]$ from $\boldsymbol{p}$ can be reached at time $T$ after resetting. It can be done by resetting the user to the corresponding direction and adjusting the translation gain to $g_t = {v\cdot T}/{s}$. To further verify the navigability for a queried position $\boldsymbol{q}$ in the PE, we test whether the straight line connecting $\boldsymbol{q}$ and the user's current position $\boldsymbol{p}$ intersects with any obstacles.

Nevertheless, $S_{near}\left(T\right)$ is not the minimum reachable displacement at time $T$. Recall that we can still shorten the reachable displacement through the curvature gain anyway. By applying the maximum curvature gain, i.e., the radius of the threshold $R$, we can then get an even smaller reachable displacement $S_{min}\left(T\right)$ by bending the straight path to its maximum extent:
\begin{equation}
S_{min}\left(T\right) = 2R\sin{\frac{v\cdot T}{g_{t}^{max}\cdot 2R}} = 2R\sin{\frac{S_{near}\left(T\right)}{2R}}
\end{equation}

$S_{min}$ is the chord length corresponding to an arc of length $S_{near}$ on a circle of radius $R$. Having a displacement $S_{min}$ means walking jointly using the maximum curvature gain and minimum translation gain in time $T$. Owing to the fact that both gains have reached their bounds, $S_{min}\left(T\right)$ is the minimum displacement that can be reached using the given time $T$.

For any position $\boldsymbol{q}$ with a displacement $s \in \left[S_{min}\left(T\right), S_{near}\left(T\right)\right)$ away from $\boldsymbol{p}$, we can steer the user to let them reach $\boldsymbol{q}$ at time $T$ by relaxing the curvature gain radius to a suitable value. Suppose the proper curvature gain radius for arriving at $\boldsymbol{q}$ is $R'$, the relationship between $R'$ and $s$ is:
\begin{equation}
\label{R'}
2R'\sin{\frac{v\cdot T}{g_{t}^{max}\cdot 2R'}}=s \quad \left(s \in \left[S_{min}\left(T\right), S_{near}\left(T\right)\right), R'\geq R\right)
\end{equation}

Equation~(\ref{R'}) is a \textbf{transcendental equation} with respect to $R'$ and there is no algebraic solution. However, we can just re-use the $k$ different $\left\{r_i\right\}$ in Sec.~\ref{3.1} to approximate the optimal solution. Thanks to the fact that the left side of Equation~(\ref{R'}) is monotonically increasing in its domain of definition, the optimal solution for $R'$ can be quickly found by binary searching among $\left\{r_i\right\}$. Once $R'$ is found, we further verify whether the arc path could be blocked by any obstacles in PE. We connect $\boldsymbol{q}$ and $\boldsymbol{p}$ with the arc of radius $R'$ and check if the arc intersects with the obstacles.

Note that adjusting the translation gain is also a potential way to reach $\boldsymbol{q}$. However this manner can't guarantee all $s \in \left[S_{min}\left(T\right), S_{near}\left(T\right)\right)$ are covered in geometry. Moreover, the user's sickness exacerbates when a larger curvature gain is applied while their walking is accelerated. Thus, we only fine-tune the curvature gain to find candidate paths in that case.

By above steps, we have found paths with less sickness to reach all possible positions from $\boldsymbol{p}$ with displacements $s \in \left[S_{min}\left(T\right), S_{max}\left(T\right)\right]$ at time $T$. We have also verified the navigability of the path from $\boldsymbol{p}$ to a specified position. We denote the set of all positions reachable from $\boldsymbol{p}$ at time $T$ as $\varphi\left(\boldsymbol{p},T\right)$.

\subsection{The Multi-User Online RDW Controller}

To coordinate the redirections of all users for a fewer total reset number, we first analyze the current state of each user in terms of the remaining time to their target and their maximum time to encounter an obstacle. We then find the bottleneck user that needs to be reset in the shortest time, and reset all the users to their reachable poses that allow them to walk for their longest possible distances once a common reset is triggered.

\textbf{Skeleton Poses \ \ }
As the number of users' poses in the physical space is infinite, we abstract the physical space with a series of discrete \emph{skeleton poses} to represent the possible states in it. We divide a given physical space $E$ into squares of side lengths $\delta$ and use the centers of the squares ${\boldsymbol{p}_{c}}$ (except those in obstacles) as the \emph{skeleton positions} $\chi\left(E\right) = \left\{\boldsymbol{p}_{c}\right\}$. 
We define $\lambda$ \emph{skeleton orientations} as $\Theta = \left\{{2\pi i}/{\lambda}\mid i=1,2,...,\lambda\right\}$. The \emph{skeleton positions} and the \emph{skeleton orientations} then form all the \emph{skeleton poses} $\Gamma\left(E\right)=\left\{\left(\boldsymbol{p},\theta\right)\mid \boldsymbol{p}\in \chi\left(E\right), \theta \in \Theta \right\}$ for the given space. 

\begin{figure}[h]
    \centering
    \includegraphics[width=0.45\linewidth]{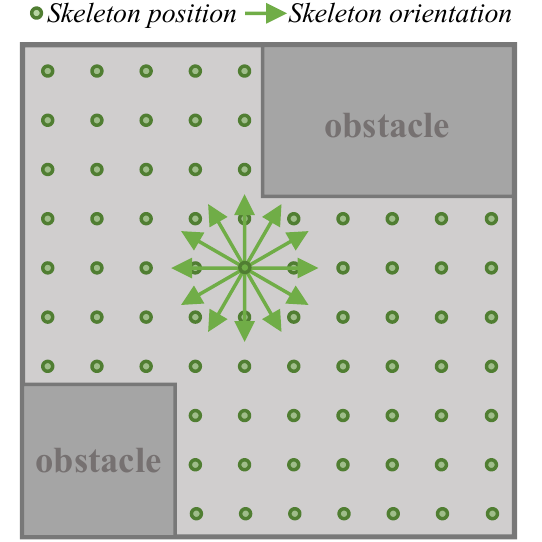}
    \caption{
    The illustration of the \emph{skeleton positions} and the \emph{skeleton orientations}.
    }
    \label{fig:skeleton}
\end{figure}

We perform an offline preprocessing by computing the maximum walking time before encountering obstacles $T_{max}\left(P\right)$ for all \emph{skeleton poses} $P \in \Gamma\left(E\right)$ . After that, we measure each \emph{skeleton position} $\boldsymbol{p}_{c}$ in $\chi\left(E\right)$ with two indicators:
\begin{align}
L\left(\boldsymbol{p}_{c}\right)&=\max \limits_{\theta \in \Theta}T_{max}\left(\left(\boldsymbol{p}_{c},\theta\right)\right)
\\
H\left(\boldsymbol{p}_{c}\right)&=\lambda({\sum \limits_{\theta \in \Theta}\frac{1}{T_{max}\left(\left(\boldsymbol{p}_{c},\theta\right)\right)}})^{-1}
\end{align}

We call $L\left(\boldsymbol{p}_{c}\right)$ the escapability of position $\boldsymbol{p}_{c}$, which represents the longest walking time a user can travel from position $\boldsymbol{p}_{c}$ if the user is just reset to the optimal direction at $\boldsymbol{p}_{c}$. We call $H\left(\boldsymbol{p}_{c}\right)$ the safety of position $\boldsymbol{p}_{c}$, which represents the harmonic mean of the maximum walking times in all \emph{skeleton orientations} of $\boldsymbol{p}_{c}$ and indicates the overall safety in case the user turns to an unknown direction at $\boldsymbol{p}_{c}$. Here we use the harmonic mean instead of the arithmetic mean to make $H\left(\boldsymbol{p}_{c}\right)$ more sensitive to the directions with a shorter maximum time to encountering an obstacle.


\textbf{User State Analysis \ \ }

The user's movement in VE is typically composed of walking toward the current virtual target and turning around to the next virtual target. One segment of immersive straight walking begins when the user have turned around or finishes a reset, and ends when the user reaches the target or a reset occurs. For the users' immersion, our RDW controller redirects each user with constant gains during one segment of their straight walking, and only activates the decision-making process for new gains when a common reset occurs or the user has turned around. 

As the user's walking speed in VE is known to the RDW controller, we can estimate the remaining time $\tau_i$ for each user $u_i$ to their virtual target $w_i$. We suppose that $u_i$ will keep walking in VE within time $\tau_i$.
Meanwhile, we calculate the maximum time for user $u_i$ to encounter obstacles in all \emph{skeleton orientations} $\theta \in \Theta$ at their current position $\boldsymbol{p}_i$, and find the largest maximum time to encounter obstacles among all \emph{skeleton orientations} as:
\begin{equation}
\Pi\left(u_i\right)=\max \limits_{\theta \in \Theta}{T_{max}\left(\left(\boldsymbol{p}_i,\theta\right)\right)}
\end{equation}

$\Pi\left(u_i\right)$ implies the upper bound of $u_i$'s walking time after being reset to the best orientation at their current position $\boldsymbol{p}_i$. If $\Pi\left(u_i\right) \geq \tau_i $, the user $u_i$ could get to the target $w_i$ without encountering obstacles again after the current reset, we then put $u_i$ into a \emph{Safe User Set} $U_S$. But if $\Pi\left(u_i\right) < \tau_i $, the user $u_i$ cannot reach his target $w_i$ after the current reset and will need a 
further reset before reaching $w_i$, we thus put him into the \emph{Unsafe User Set} $U_N$.

\textbf{Redirection Strategy \ \ }
For each user $u_i$, we have three essential ways to reset and redirect them, i.e. reset and redirect the user to walk for the maximum permitted time, walk to the position with the largest escapability or walk to the position with the largest safety. 

If $U_N$ is not empty, the ``unsafe" users will need at least one more reset before reaching their targets, as they couldn't get to their targets in one fell swoop no matter what direction they are reset to. In this case, the user in $U_N$ with the smallest maximum time to encounter obstacles is supposed to cause the next reset for other users, so we consider that user as the ``bottleneck" user. We denote the subscript of the bottleneck user as $*$, so:
\begin{equation}
u_*=\arg \min \limits_{u_i \in U_N}{\Pi\left(u_i\right)}
\end{equation}

We reset and redirect the bottleneck user $u_*$ to walk for the exact maximum permitted time $\Pi\left(u_*\right)$ to maximize the time interval to the next common reset for all the online users. In order to do that, we reset $u_{*}$ to the optimal direction of $\arg \max \limits_{\theta \in \Theta}{T_{max}\left(\left(\boldsymbol{p}_{*},\theta\right)\right)}$, and steer $u_*$ with the curvature gain leading to the longest walking time $\Pi\left(u_*\right)$, where the gain is selected from the $\left(2k+1\right)$ candidate curvature gains mentioned in Section \ref{3.1}. 

For other ``safe" or ``unsafe" users whose remaining time are greater than the bottleneck time, since they are expected to reset at time $\Pi\left(u_{*}\right)$, we need to reset and redirect them to more favorable physical positions in time $\Pi\left(u_{*}\right)$ so that they can travel further after the reset triggered by the bottleneck user $u_{*}$. To achieve this goal, we estimate the reachable spatial extent after reset $\varphi\left(\boldsymbol{p_i},\Pi\left(u_{*}\right)\right)$ using the algorithm in Section \ref{3.2} within the bottleneck time $\Pi\left(u_{*}\right)$ for each of the other users. For a specific user $u_i$, if there are \emph{skeleton positions} in their reachable area $\varphi\left(\boldsymbol{p_i},\Pi\left(u_{*}\right)\right)$, we select the one of all the reachable \emph{skeleton positions} with the largest escapability $L\left(\boldsymbol{p}_{c}\right)$ as their physical destination, denoting as $\boldsymbol{p}^{D}_{i}$, and redirect the user $u_i$ to $\boldsymbol{p}^{D}_{i}$ using the reset direction and gains obtained using the method mentioned in \ref{3.2}. 
\begin{equation}
\boldsymbol{p}^{D}_{i}=\arg \max \limits_{\boldsymbol{p}_{c} \in \chi\left(E_i\right)\cap\varphi\left(\boldsymbol{p_i},\Pi\left(u_{*}\right)\right)}{L\left(\boldsymbol{p}_{c}\right)}
\end{equation}

This ensures $u_i$ to have a longer walking time after the scheduled reset to be triggered by the bottleneck user. In case there is no \emph{skeleton position} in $u_{i}$'s reachable area $\varphi\left(\boldsymbol{p_i},\Pi\left(u_{*}\right)\right)$, which is a rare case that only may happen when $\Pi\left(u_*\right)$ is too small or the \emph{skeleton positions} are too sparse, we reset and redirect $u_{i}$ to walk for the longest time $\Pi\left(u_i\right)$ instead, with the same operation as $u_{*}$.

However, for the ``safe" users with the remaining time $\le$ the bottleneck time, as they will reach their virtual targets before the common reset and turn around to the unknown new targets, it's unreasonable to also navigate them to the maximum escapability destinations with the bottleneck time. When $U_N$ is empty, since the bottleneck user doesn't exist and the bottleneck time is equivalent to $+\infty$, all users are ``safe" users with the remaining time $\le$ the bottleneck time. So such users cannot be ignored. For each such user $u_i$, we just need to reset and redirect the user to a more reassuring position for the unknown next target during the user's walking time $\tau_i$. In order to do that, we firstly calculate the reachable spatial area $\varphi\left(\boldsymbol{p_i},\tau_i\right)$ with remaining time $\tau_i$ for the user $u_i$. Then we find the \emph{skeleton positions} in $\varphi\left(\boldsymbol{p_i},\tau_i\right)$ with the greatest safety $H\left(\boldsymbol{p}_{c}\right)$ as $u_i$'s physical destination $\boldsymbol{p}^{D}_{i}$, thereby to determine the reset direction and redirection gains.
\begin{equation}
\boldsymbol{p}^{D}_{i}=\arg \max \limits_{\boldsymbol{p}_{c} \in \chi\left(E_i\right)\cap\varphi\left(\boldsymbol{p_i},\tau_i\right)}{H\left(\boldsymbol{p}_{c}\right)}
\end{equation}

The reason we optimize the position by checking the overall safety $H\left(\boldsymbol{p}_{c}\right)$ rather than the escapability $L\left(\boldsymbol{p}_{c}\right)$ is to cope with the user's unpredictable turning direction to the next random target. Once $u_i$ has reached his current target $w_i$ and has turned ready for moving to the next target, we calculate the longest time he can walk from his new pose, and redirect $u_i$ accordingly until a new reset is triggered by some user. If a common reset is triggered just when the user is turning, we allow the user to continue to turn in place but not move until the common reset ends. If the user starts walking before the common reset is completed, the resetting interface will pop up to guide the user back to the starting position. When everyone is done, all users can start moving at the same time.


\section{Experiments}
\subsection{Implementation Details}
In our experiments, we use $k=10$ as the number of different candidate radii for the time span and spatial extent estimation. Therefore, considering the steering direction and zero curvature gain, there are a total of $21$ potential curvature gains. We set the spacing between the \emph{skeleton poses} as $\delta=0.5m$, which is a suitable density for users' possible position since the adult's shoulder width is ($34cm\sim 38cm$). The number of orientations of skeleton poses is set to $\lambda=30$ for a dense scan of the potential directions. The radius of curvature gain threshold is set to $R=7.5m$, and the translation gain thresholds are set to $g_{t}^{min}=0.86$ and $g_{t}^{max}=1.26$, respectively. The gain thresholds have been verified and commonly used by extensive RDW literature~\cite{azmandian2015physical,bachmann2013collision,hodgson2013comparing,thomas2019general,lee2020optimal,bachmann2019multi,messinger2019effects,dong2020dynamic,dong2021dynamic,williams2021arc,williams2021redirected}.
With the above parameters, our RDW controller runs in real-time on a PC with Intel Core i7-9750H CPU 2.60GHz and 8GB of RAM.

Simulation is widely recognized as an effective way to evaluate RDW methods~\cite{bachmann2013collision,azmandian2015physical,azmandian2017evaluation,thomas2019general,messinger2019effects,dong2020dynamic,strauss2020steering,williams2021arc,williams2021redirected}. 
We use the motion model proposed by Azmandian \etal~\cite{azmandian2015physical} to simulate the users' movements in VE. As with most methods, we let the simulated user in VE translate at a constant speed of $1$m/s and rotate at a constant rate of $\pi/2$ rad/s. The user's starting position is random in PE. A simulation trial ends when all users have walked more than $400$ meters, and we have $100$ simulate trials for each method under each PE configuration.



We compare our method with several state-of-the-art RDW methods, including Steer-to-Center(S2C)~\cite{hodgson2013comparing}, Steer-to-Orbit(S2O)~\cite{hodgson2013comparing}, ZigZag~\cite{razzaque2005redirected}, Dynamic APF (DAPF)~\cite{dong2020dynamic} and the steer-by-reinforcement learning method (SRL)~\cite{strauss2020steering}. 
To make the comparison as fair as possible, when a common reset happens in the comparison methods, we do not make the users who were not encountering obstacles wait in vain, but reset them simultaneously to better orientations of their current locations. In addition, we use reset-to-gradient(R2G)~\cite{thomas2019general} as the reset strategy for the methods (S2C,S2O and Zigzag) without a reset policy.


\begin{figure*}[t]
    \centering
    \includegraphics[width=0.95\linewidth]{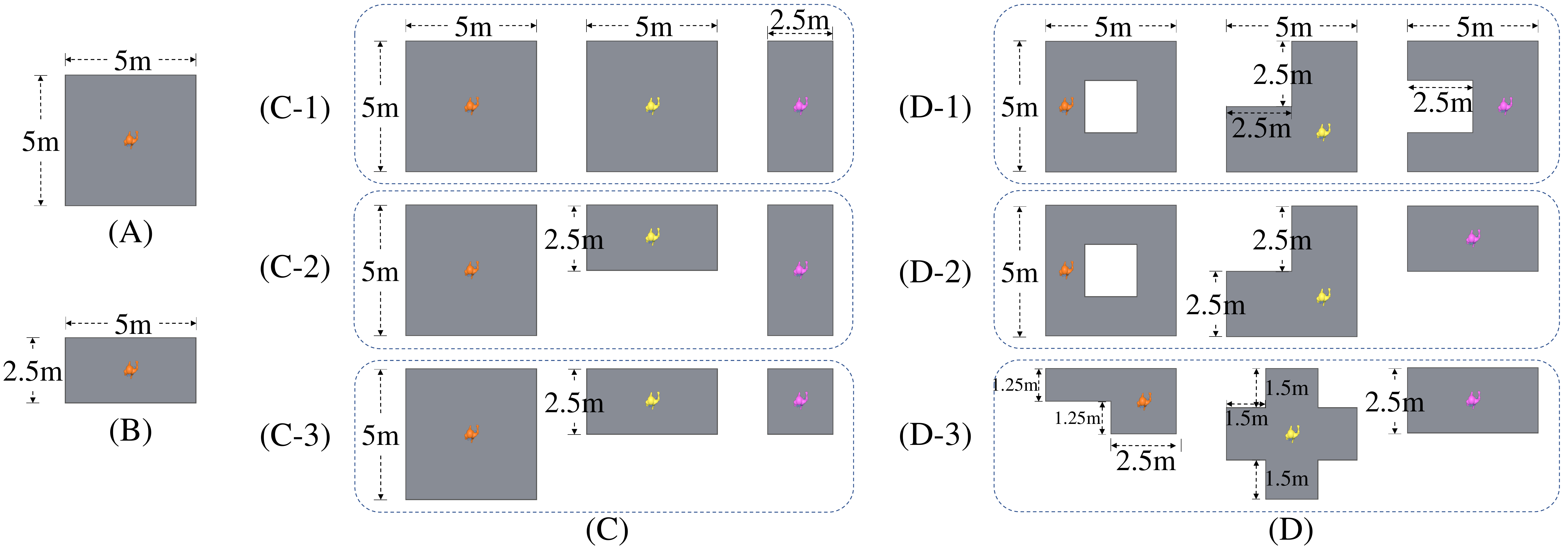}
    \caption{The physical environment layouts used for experiments.
    }
    \label{fig:PElayouts}
\end{figure*}

\subsection{Performance with Different Numbers of Users}

Intuitively, the more online users, the more frequently users trigger resets. Therefore, we firstly evaluate how the user quantity affects the number of resets for our method and the comparison methods. To control for variables, we use identical PE for each user and test  methods with varying numbers of users. Specifically, we first configured a square PE with a side length of $5$m for each user as shown in Fig.~\ref{fig:PElayouts} (A), and tested the methods with different user numbers of $2$, $3$, $4$ and $6$.

\begin{figure*}[h]
    \centering
    \includegraphics[width=0.99\linewidth]{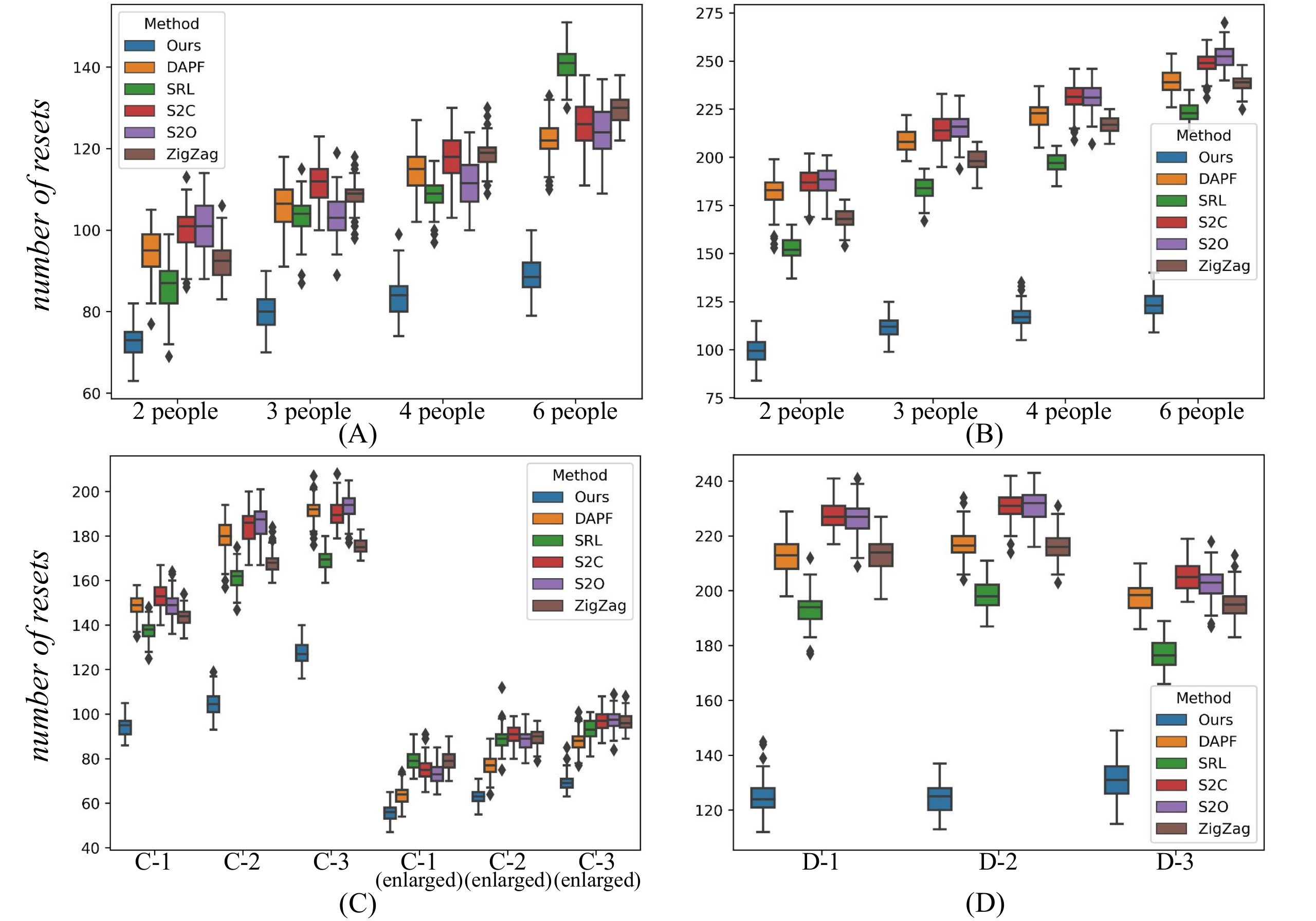}
    \caption{
    The box-plots of the number of resets. (A) is the box-plot for the results of each method tested with different numbers of users located in off-site indentical $5m$x$5m$ square PEs. (B) is the box-plot for the results of each method tested with different numbers of users located in off-site indentical $2.5m$x$5m$ rectangle PEs. (C) is the box-plot for the results of each method tested with $3$ users located in off-site unequal PEs. (D) is the box-plot for the results of each method tested with $3$ users located in off-site unequal and irregular PEs.
    }
    \label{fig:simulation}
\end{figure*}

The distributions of the number of resets for all methods under different user quantities are shown in the box-plots in Fig.\ref{fig:simulation} (A). We can find from the distribution that the number of resets for our method is always significantly smaller than that for the comparison methods regardless of the online user quantity. We can also find that the number of resets for all methods increases as the online user quantity increases, but the increase in the reset times for our method is noticeably slower than that for the comparison methods.

Although we reset all the users to their optimal orientations to reduce the resets for the comparison methods, the comparison methods do not have the scheme to consider the collaboration between different users, which makes users often implicated in the reset of other users. On the contrary, our method reasons and maximizes the bottleneck time to the subsequent common reset for all the users and redirects the users to more favorable positions in their PEs to prepare for the next reset in advance, thus triggering the fewest common resets.

The numbers of resets of the comparison methods increase rapidly along with the user number. Since the comparison methods do not coordinate the redirection of all users, it is not surprising that the probability of being implicated in a reset triggered by other user will boost for each user as the user number increases. As our method maximizes the time interval to the next common reset for all the online users, our method is more advantageous when the number of users increases.

To further increase the challenge of the evaluation, we next upgraded the difficulty of the PE by switching it to an even smaller shape, i.e. a rectangle with one side only half the length of the original, as shown in Fig.~\ref{fig:PElayouts} (B). We also tested the performance of all methods with different user numbers of $2$, $3$, $4$ and $6$, and the results are shown in Fig.\ref{fig:simulation} (B). As resets are more easily to be triggered in such small PEs, from the results we can see that the number of resets for all methods becomes larger in this kind of more challenging environment, but our method demonstrates more advantages over the other methods.

We conducted the Kruskal-Wallis H test on all the above results and found that the result distributions are statistically different. Post hoc pairwise comparisons by Mann–Whitney U test find satisfactory significance between our method and all the comparison methods ($p<<0.01$ after Bonferroni adjustment).


Despite the statistical differences between the comparative methods and our method, we want to clarify that the the comparison methods were not designed for off-site multiplayer online VR scenarios, so the results do not mean that the comparison methods are poor. We also find from the results that S2C may not be a good idea for the off-site multiplayer online VR scenarios compared to S2O. If the user is too close to the PE center at the time when a common reset occurs, the user may not have a maximum walking distance after the reset, so the next reset will be triggered earlier. This makes it possible that S2C may be inferior to methods such as S2O in the off-site multiplayer online VR cases, although S2C is proved to perform better than S2O in most general cases~\cite{azmandian2015physical}. 
\subsection{Performance on Unequal and Irregular PEs}

Since real-world online VR users are usually distributed in physical spaces of different shapes, we then test the performance of all the above methods under unequal PE layouts to better evaluate the method in real-world use cases. To control for variables, we fix the number of online users to $3$ in these experiments, and use various combinations of unequal PE layouts to test the RDW controllers. Specifically, we use the three combinations of layouts shown in Fig.~\ref{fig:PElayouts} (C) to test the methods. In these three combinations of PE layouts, part of the PEs become smaller than others, and the combination gets harder and harder from (C-1) to (C-3). In addition to the layout shape, the PE sizes also affect the number of resets. To further evaluate the impact of PE size on methods' performance, we reuse the PE layout combinations in Fig.~\ref{fig:PElayouts} (C) and enlarge each PE by doubling its original side length to make a larger PE with the same shape, and then we test these methods again in the larger PEs. So totally we have tested the methods under 6 sets of PE combinations.

The distributions of the results are reported in Fig.\ref{fig:simulation}(C) as box-plots. From the results we can find that our method has an absolute advantage over the comparison methods when the PE layouts are unequal. That is because in unequal PEs the users with relatively small spaces will easily become ``bottleneck" users and can markedly ``drive" other users to reset. However, by analyzing the bottleneck time to the next reset and redirecting all the users accordingly, our method can synchronize the resets for more users and make a larger time interval between common resets. Although the number of resets get smaller in the enlarged unequal PEs for all the methods, the Kruskal-Wallis H test and the post hoc comparisons by Mann–Whitney U test still confirmed the statistical difference between our method and the comparison methods ($p<<0.01$ after Bonferroni adjustment).

Since the PE is usually irregular due to obstacles in real-world scenarios, we then challenge to test the methods with unequal and irregular PEs. We add various obstacles into the PE to make the PE irregular, and the irregular PE layout combinations used for testing are shown in Fig.~\ref{fig:PElayouts} (D).The number of users is also fixed to $3$. As we can see, the irregular PEs are smaller, usually non-convex and sometimes have voids inside due to the obstacles.

The box plots of the number of resets for irregular PE combinations are shown in Fig.\ref{fig:simulation}(D). As complex obstacles limit the user's walking, the reset times for all methods are highly increased. However, we can find that our method still significantly improves over the comparison methods in these challenging irregular PEs. This illustrates that our mechanism for maximizing the time interval to the next common reset can work effectively in the irregular PEs to greatly reduce the number of resets.

\section{User Study}
\subsection{Experiment Design}
In order to assess our method in real-world applications, we conducted a live user study. We implemented our algorithm with VIVE Pro Eye headsets in Unity, and conducted experiments with a \textbf{three-user} scenario.
We assign the users to three separate physical spaces, each $5m*3m$, similar to common household rooms. During the experiment, each participant with a headset was located in their own physical space, exploring a shared VE simultaneously. We mounted two tracking stations on two opposite sides for each space, ensuring that each user in their venue could be tracked. We use the Photon Unity Networking 2 to implement the communication among the three sets of equipment. Each participant starts at their site center, facing north. In our experimental online VR application, each user is required to walk toward a randomly generated target in the virtual space. Once the participant reaches their target, a new target will be generated. As a fair online application, when one of the participants has to be reset, the RDW controller makes all of the users reset at the same time. Considering that the position may also change during the reset, we included a position calibration function for the reset interface: by having the user align two arrows in the reset interface, we ensure the users return to the original physical position where they were prompted to reset. It is difficult for our test to record the number of resets with a specified physical walking distance since the three participants may have different walking distances. Alternatively, we recorded the distance each user has traveled in VE when a fixed number of $8$ common resets have been triggered as the measurement to evaluate all the RDW methods. 
In the reported results below, \textbf{the longer the users walked, the better the RDW method is}. We use the state-of-the-art method DAPF~\cite{dong2020dynamic} as the comparison method in the user study. As the participants' walking speed $v$ is unknown in advance, we instructed the participants to walk calmly as in ~\cite{razzaque2001redirected}, and used the average speed $v$ of the last segment of their straight walking as the estimated walking speed of the following segment of straight walking.

\subsection{Results}
\label{user_study_result}
\begin{figure}[h]
    \centering
    \includegraphics[width=1.0\linewidth]{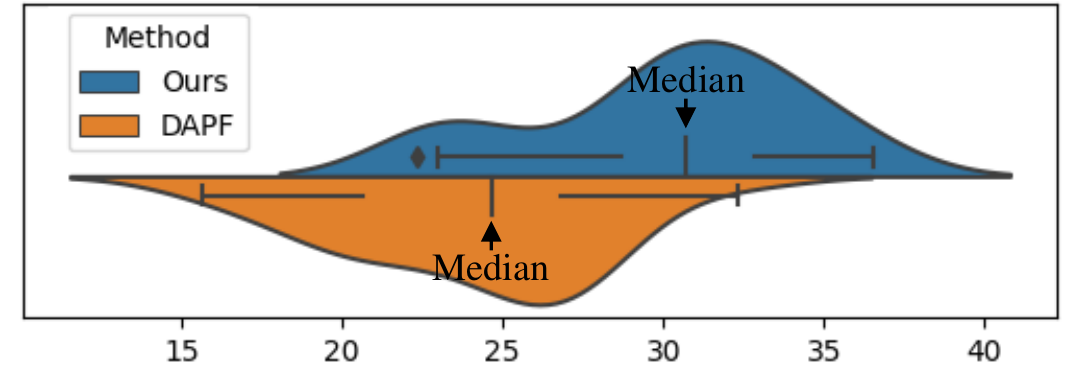}
    \caption{
    The distributions of the participants' average walking distance (m).
    }
    \label{fig:user_study_result}
\end{figure}

We recruited $27$ participants (10 females and 17 males, aged 19 to 28) to join our tests. Since one experiment requires three participants, we randomly assigned the participants to different groups and asked each of them joined three experiments. Thus we conducted a total of $27$ multi-user experiments. In each experiment, the 3 participants were given adequate rest time between the two trials of the comparison methods. The results are shown in Fig.~\ref{fig:user_study_result}.
From Fig.~\ref{fig:user_study_result}, we can find that users can have a larger expected walking distance between resets using our method, which results in a better immersive experience. Mann-Whitney U test further finds the statistical difference between the comparison methods ($U=109.0,P<.001$).
We had also asked the participants to fill an SSQ~\cite{kennedy1992simulator} table to record their subjective feelings before and after the trials in terms of the quality of the experiences. The SSQ records find no significant difference between our method and DAPF, showing that the participants were not feeling noticeably uncomfortable with either method.

\section{Limitations and future work}

Our work has several limitations. We only use a fixed curvature gain during one segment of the user's walk before stopping in physical space. However, if we do not consider the sacrifice of the user comfort, we can steer the user with dynamically changing curvature gains within one segment of walk to make the user's physical path more flexible, thus the actual walking time to a given physical position will also be longer.  

As the user's walking time is naturally related to the walking speed in VE, our calculations for both the time before encountering obstacles and the reachable area at a given time are related to the users' speeds. 
Although most existing methods assume the users to have constant walking speeds, this assumption may not hold if the user walks too fast in VE, as the speed can fluctuate greatly. A more desirable speed estimation strategy would help our method cope with this situation.

Although our method has outperformed the state-of-the-art methods in online VR scenarios, the performance of our method could be further improved by involving rotation gains. By adjusting rotation gains, we will be able to redirect users to a better orientation with a more refined control when they turn to walk towards new virtual targets, which could further increase the user's longest walking time and reduce the overallnumber of resets. The strategy considering rotation gains remains a future work of our method.

\section{Conclusion}
In this work, we propose a redirected walking method for online multiplayer VR applications, which coordinates the redirections and resets for all the users in different physical spaces to reduce the total reset number. We achieve our goal by first estimating all the users' walking time before encountering obstacles and reachable areas to find the bottleneck user leading to the next reset. Then, we redirect all users to their favorable poses before resetting. The simulation experiments and user study
demonstrate the ability of our method to provide more continuous walking experiences for multiple online players in the same virtual environment. 

\ifCLASSOPTIONcompsoc
  \section*{Acknowledgments}
\else
  \section*{Acknowledgment}
\fi

This work was supported by the National Natural Science Foundation of China (Project Number 62132012), Tsinghua-Tencent Joint Laboratory for Internet Innovation Technology and Marsden Fund Council managed by Royal Society of New Zealand (No. MFP-20-VUW-180).

\ifCLASSOPTIONcaptionsoff
  \newpage
\fi



%

\bibliographystyle{IEEEtran}
\bibliography{reference.bib}

%

\begin{IEEEbiography}[{\includegraphics[width=1in,height=1.25in,clip,keepaspectratio]{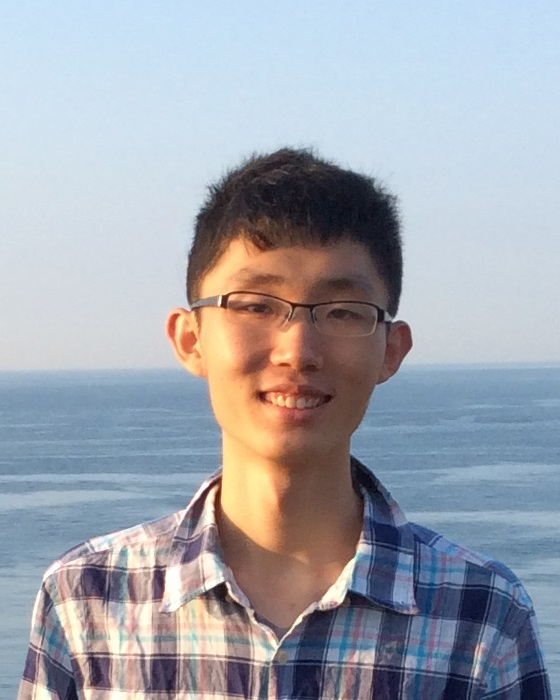}}]{Sen-Zhe Xu} received the PhD degree of Computer Science and Technology from Tsinghua University, Beijing, in 2021. He is currently a postdoc in Yau Mathematical Sciences Center at Tsinghua University. His research interests include virtual reality and image/video processing.
\end{IEEEbiography}

\begin{IEEEbiography}[{\includegraphics[width=1in,height=1.25in,clip,keepaspectratio]{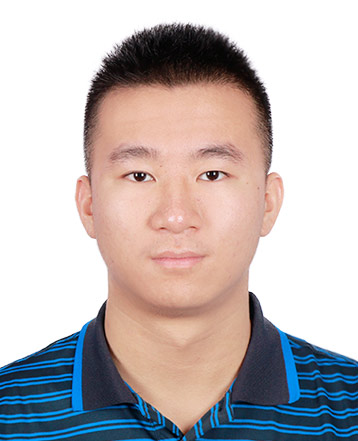}}]{Jia-Hong Liu} is an undergraduate student in the Department of Computer Science and Technology at Tsinghua University. His research interests include virtual reality and 3D scene reconstruction.
\end{IEEEbiography}


\begin{IEEEbiography}[{\includegraphics[width=1in,height=1.25in,clip,keepaspectratio]{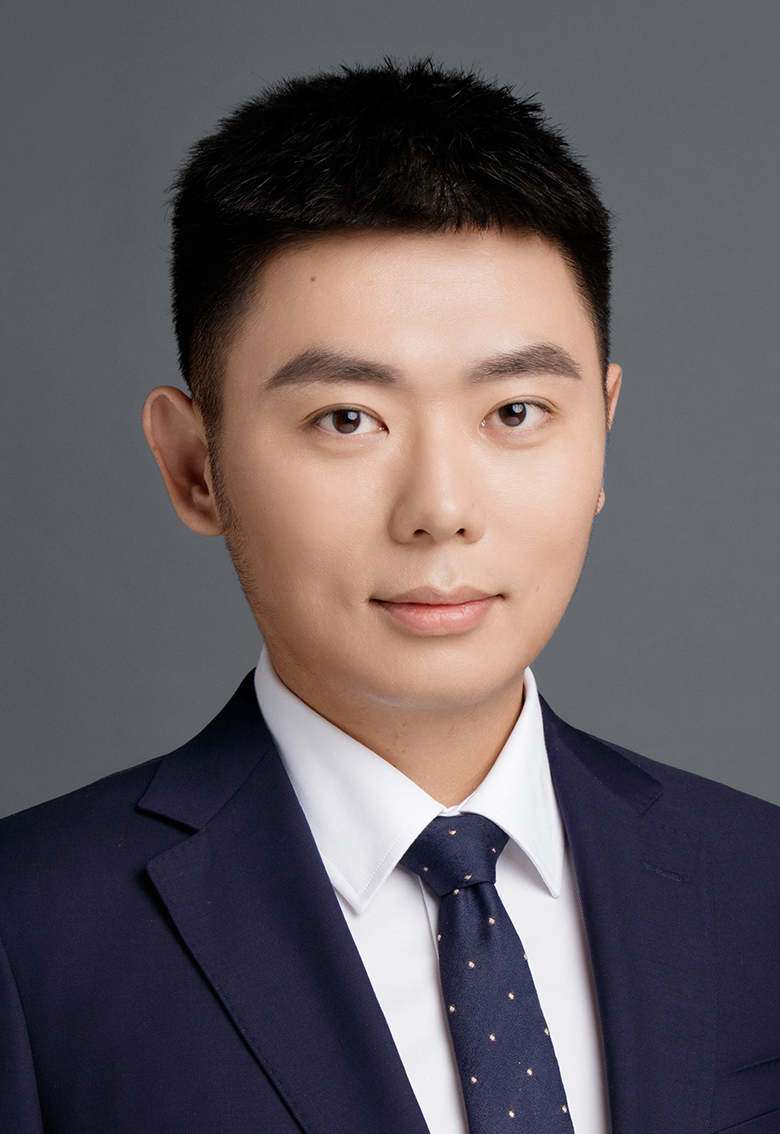}}]{Miao Wang}
is an Associate Professor with the State Key Laboratory of Virtual Reality Technology and Systems, School of Computer Science and Engineering, Beihang University, China. His research interests include Virtual Reality, Computer Graphics and Visual Computing. During 2016-2018, he did postdoc research in Visual Computing at Tsinghua University. He received his Ph.D. degree from Tsinghua University in 2016 and the bachelor degree in Xidian University in 2011. He serves a program committee member of IEEE VR and ISMAR conferences. He is a member of IEEE, ACM and Asiagraphics.
\end{IEEEbiography}

\begin{IEEEbiography}[{\includegraphics[width=1in,height=1.25in,clip,keepaspectratio]{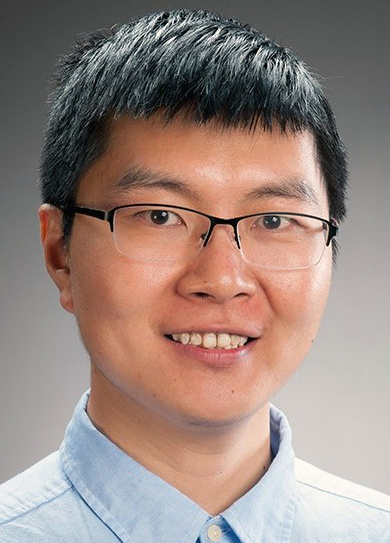}}]{Fang-Lue Zhang} is currently a lecturer with Victoria University of
Wellington, New Zealand. He received the Bachelors degree from Zhejiang University, Hangzhou, China, in 2009, and the Doctoral degree from Tsinghua University, Beijing, China, in 2015. His research interests include image and video editing, computer vision, and computer graphics. He is a member of IEEE and ACM. He received Victoria Early Career Research Excellence Award in 2019. He is a committee member of IEEE Central New Zealand sector.
\end{IEEEbiography}

\begin{IEEEbiography}[{\includegraphics[width=1in,height=1.25in,clip,keepaspectratio]{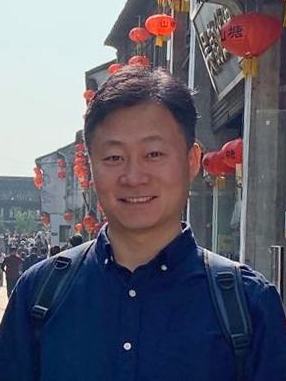}}]{Song-Hai Zhang} received the PhD degree of Computer Science and Technology from Tsinghua University, Beijing, in 2007. He is currently an associate professor in the Department of Computer Science and Technology at Tsinghua University. His research interests include computer graphics, virtual reality and image/video processing.
\end{IEEEbiography}





\end{document}